\documentclass[prd,aps,twocolumn,showpacs, nofootinbib,superscriptaddress,notitlepage]{revtex4-1}
\usepackage{amsmath,graphicx,epsfig,amssymb}
\usepackage{inputenc}
\usepackage[usenames]{color}
\usepackage{slashed}
\usepackage{multirow}
\usepackage{subfigure}
\usepackage{dsfont}
\usepackage{comment}
\usepackage{steinmetz}
\usepackage{setspace}
\usepackage{enumerate}

\allowdisplaybreaks

\usepackage{floatrow}
\newfloatcommand{capbtabbox}{table}[][\FBwidth]

\begin{document}

\title{Heavy quark mass dependence of the $\Lambda_Q$ light-cone distribution amplitude in QCD}

\author{Yu-Ji Shi}
\email{shiyuji@ecust.edu.cn (corresponding author)}
\affiliation{School of Physics, East China University of Science and Technology, Meilong Road 130, Shanghai 200237, China}

\author{Ji Xu}
\email{xuji@lzu.edu.cn (corresponding author)}
\affiliation{Frontiers Science Center for Rare Isotopes, and School of Nuclear Science and Technology, Lanzhou University, Lanzhou 730000, China}

\author{Shuai Zhao}
\email{zhaos@tju.edu.cn}
\affiliation{Department of Physics and Center for Joint Quantum Studies,
School of Science, Tianjin University, Tianjin 300350, China}

\begin{abstract}
We study the heavy quark mass dependence of the leading-twist light-cone distribution amplitude (LCDA) of the $\Lambda_Q$ baryon in QCD. 
Starting from the factorization formula that relates the QCD LCDA to the boosted heavy-quark effective theory (bHQET) LCDA, we derive a first-order partial differential equation governing this mass dependence in the peak region. 
The equation is solved analytically, and the explicit factor connecting LCDAs at different heavy quark masses is presented. 
We further incorporate Borel‑resummed perturbative corrections from a renormalon model into the factorization.
The impact of these renormalon corrections on the mass dependence is studied, and a numerical analysis using a simple LCDA model is performed to illustrate the behavior and to assess the uncertainties arising from the corrections, thereby providing a numerical estimate of the associated power corrections to the mass dependence. 
Our results provide an essential tool for extrapolating lattice QCD calculations of heavy-baryon LCDAs from smaller simulated masses to the physical bottom quark mass.
\end{abstract}
\date{\today}

\maketitle

\section{Introduction}
In recent years, the search for direct $CP$ violation (CPV) has made weak decays of heavy baryons an important topic in flavor physics. A major breakthrough came in 2025, when the LHCb collaboration reported evidence for CPV in $\Lambda_b^0 \to pK^-\pi^+\pi^-$~\cite{LHCb:2025ray}, a decay mode for which theoretical studies had previously predicted CPV~\cite{Wang:2024oyi,Han:2024kgz}.
Predictions for $\Lambda_b$ decays generally rely on factorization-based methods such as QCD factorization (QCDF)~\cite{Beneke:1999br,Beneke:2000ry}, soft-collinear effective theory (SCET)~\cite{Bauer:2000yr,Bauer:2001yt,Bauer:2002nz,Wang:2011uv,Lu:2025gjt}, perturbative QCD (PQCD)~\cite{Shih:1998pb,Keum:2000wi,Lu:2000em,Keum:2000ph,Lu:2009cm,Han:2022srw,Han:2025tvc,Li:2025rsm,Yang:2025yaw}, and light-cone sum rules (LCSR)~\cite{Wang:2009hra,Wang:2015ndk,Miao:2022bga,Khodjamirian:2023wol,Huang:2024oik}. These methods separate the short-distance perturbative effects from the long-distance non-perturbative effects. In this framework, the $\Lambda_b$ light-cone distribution amplitudes (LCDAs) serve as the key non-perturbative inputs. The LCDAs describe the parton structure of the baryon and directly enter the calculation of decay amplitudes. Therefore, precise knowledge of the $\Lambda_b$ LCDAs is essential for making reliable predictions in heavy baryon decays.

Current determinations of $\Lambda_b$ LCDAs rely mainly on QCD sum rules~\cite{Ali:2012pn} and model parameterizations ~\cite{Bell:2013tfa,Feldmann:2025dcs}.  Beyond these model-based approaches, a first-principles lattice QCD method is highly desirable. However, such a lattice calculation faces two main theoretical challenges. First, the light-cone definition prohibits direct lattice calculation. Second, the HQET formulation with the effective heavy-quark field $h_v$ introduces additional technical difficulties.
Recently, a two-step matching scheme has successfully addressed both challenges for heavy-meson LCDAs~\cite{Han:2024fkr,LatticeParton:2024zko,LPC:2026vyv} and can be applied equally to heavy baryons. This framework involves two sequential effective-theory steps. In the first step, factorization based on the Large-Momentum Effective Theory (LaMET)~\cite{Ji:2013dva,Ji:2014gla,Ji:2020ect,Cichy:2018mum,Xiong:2026fvs} maps the equal-time quasi-DA of the heavy baryon onto its light-cone counterpart, the QCD LCDA, at scales below the large momentum $P^z$ of baryon. In the second step, collinear modes with off-shellness $m_Q^2$ are integrated out, matching the QCD LCDA onto its HQET counterpart in a boosted frame. This step has been accomplished at leading power for both the heavy meson and the $\Lambda_b$ baryon~\cite{Ishaq:2019dst,Zhao:2019elu,Beneke:2023nmj,Shi:2026mjb}.

The QCD LCDAs of a heavy baryon involves two scales: the renormalization scale $\mu$ and the heavy-quark mass $m_Q$. Its evolution with respect to the renormalization scale can be expressed in terms of the two-particle kernels familiar from the Lange-Neubert (LN) evolution equation of the B-meson LCDA~\cite{Lange:2003ff,Ball:2008fw}. In contrast, its dependence on $m_Q$ has not been systematically explored.
Understanding this mass dependence is of particular importance for lattice QCD calculations. Limited by computational resources, lattice simulations can typically acess only heavy-quark masses that are lower than those in physical processes. A precise characterization of the $m_Q$-dependence of the QCD LCDA would therefore allow one to extrapolate from the simulated lower masses to the physically relevant heavier masses, which are challenging to simulate directly. For heavy mesons, such a mass evolution equation has been derived in Ref.~\cite{Wang:2024wwa}, providing a solution that relates the QCD LCDA of a heavy meson at different heavy-quark masses.

In this work, we extend the mass evolution formalism to the $\Lambda_Q$ baryon in the peak region.  
Going beyond the leading-power analysis, we incorporate Borel‑resummed perturbative 
corrections from a renormalon model~\cite{Beneke:1998ui} into the factorization between 
the QCD and HQET LCDAs~\cite{Guo:2025obm}.  These corrections allow us to estimate the missing power 
corrections, which are expected to be relevant at moderate quark masses.  
The rest of the paper is organized as follows.  
Section~\ref{sec:def1} presents the definitions of the $\Lambda_Q$ LCDAs and the
factorization formula between QCD and HQET.  
Section~\ref{sec:mass} derives the leading-power mass dependence relation of the QCD LCDA.  
Section~\ref{sec:power} presents the renormalon-model corrections and estimates the 
resulting power corrections to the mass evolution.  
Numerical results are presented in Sec.~\ref{sec:num}, and our conclusions are summarized 
in Sec.~\ref{sec:concl}.

\section{Factorization}\label{sec:def1}

In the boosted frame, the leading-twist QCD LCDA of $\Lambda_Q$ is defined by the matrix element~\cite{Shi:2026mjb}
\begin{align}
\Phi_{\rm QCD}(x_1,x_2) f_{\Lambda_Q} u_{\Lambda_Q}(p) = \langle 0 | O_c(x_1,x_2) | \Lambda_Q(p) \rangle \,,
\label{eq:defQCD}
\end{align}
where the nonlocal operator $O_c$ reads
\begin{align}
&O_c(x_1,x_2) = (p^+)^2 \! \int\!\frac{dt_1 dt_2}{(2\pi)^2}\, e^{i x_1 p^+ t_1 + i x_2 p^+ t_2}  \epsilon_{ijk}  \nonumber\\
\times & W_{ii'}(t_1n) u_{i'}^T(t_1n) \Gamma W_{jj'}(t_2n) d_{j'}(t_2n) W_{kk'}(0) Q_{k'}(0) \,,
\label{eq:Oc}
\end{align}
with $\Gamma = C\gamma_5 \slashed n$, $n^\mu = (1,0,0,-1)/\sqrt{2}$, and $W(t_i n)$ the light-like Wilson line:
\begin{align}
W(t n)={\cal P}\,{\rm exp}\left[ig_s\int_{t}^{\infty}d\lambda\  n\cdot A^a(\lambda n) T^a\right] \,.
\end{align}
The decay constant $f_{\Lambda_Q}$ is defined from the local limit of the matrix element in Eq.\,\eqref{eq:Oc}.

In the boosted frame, the corresponding HQET LCDA of $\Lambda_Q$ is defined by
\begin{align}
\Phi_{\rm bHQET}(\omega_1,\omega_2) \bar{f}_{\Lambda_Q} u_{\Lambda_Q}(p) = \langle 0 | O_h(\omega_1,\omega_2) | \Lambda_Q(p) \rangle \,,
\label{eq:defHQET}
\end{align}
with the nonlocal operator
\begin{align}
& O_h(\omega_1,\omega_2) = (v^+)^2 \! \int\!\frac{dt_1 dt_2}{(2\pi)^2}\, e^{i \omega_1 v^+ t_1 + i \omega_2 v^+ t_2} \epsilon_{ijk} \nonumber\\
 \times & W_{ii'}(t_1n) u_{i'}^T(t_1n) \Gamma W_{jj'}(t_2n) d_{j'}(t_2n) W_{kk'}(0) h_{n,k'}(0) \,,
\label{eq:Oh}
\end{align}
where $h_n$ is the large component of the heavy quark field in bHQET, $v^\mu$ the baryon velocity, and $\omega_i$ the longitudinal light-quark momenta. The decay constant $\bar{f}_{\Lambda_Q}$ is defined from the local limit of the matrix element in Eq.\,\eqref{eq:Oh}.

In the peak region $x_i \ll 1$, the QCD LCDA factorizes into a convolution of a perturbative jet function and the bHQET LCDA. To leading power in $\Lambda_{\rm QCD}/m_Q$, the matching relation takes the simple multiplicative form~\cite{Shi:2026mjb}
\begin{align}
&\Phi_{\rm QCD}(x_1,x_2) \nonumber\\
=&\  m_Q^2\, \mathcal{J}(m_Q,\mu)\, \Phi_{\rm bHQET}(x_1 m_Q, x_2 m_Q) \,,
\label{eq:FacFormula}
\end{align}
where the matching kernel has been computed up to one loop:
\begin{align}
& \mathcal{J}^{(1)}(m_Q,\mu) \equiv \frac{\bar{f}_{\Lambda_Q}}{f_{\Lambda_Q}} J_{\rm peak}^{(1)}(m_Q^2,\mu)\nonumber\\
= &1 + \frac{\alpha_s(\mu)C_F}{4\pi} \left( \frac{1}{2} \ln^2\!\frac{\mu^2}{m_Q^2} + \frac{5}{2} \ln\!\frac{\mu^2}{m_Q^2} + \frac{\pi^2}{12} + 5 \right) \,.
\label{eq:Jfunction}
\end{align}
Here $J_{\rm peak}(m_Q^2,\mu)$ is obtained by matching the nonlocal operators $O_c(x_1,x_2)$ and $O_h(\omega_1,\omega_2)$ in the peak region, while $\bar{f}_{\Lambda_Q}/f_{\Lambda_Q}$ is obtained by matching these two operators in the local limit.
Eq.\,\eqref{eq:FacFormula} is the starting point of the analysis of mass dependence relation; it relates the two LCDAs at a given renormalization scale $\mu$ and heavy-quark mass $m_Q$.

\section{Mass dependence of the QCD LCDA}
\label{sec:mass}

We now study how $\Phi_{\rm QCD}$ changes when the heavy-quark mass $m_Q$ is varied, while keeping $\mu$ fixed. Differentiating Eq.\,\eqref{eq:FacFormula} with respect to $\ln m_Q$ and using the chain rule yields
\begin{align}
&m_Q\frac{\partial}{\partial m_Q}\Phi_{\rm QCD} \nonumber\\
=\ & 2m_Q^2\mathcal{J}\,\Phi_{\rm bHQET} + \left(m_Q\frac{d\mathcal{J}}{dm_Q}\right) m_Q^2\Phi_{\rm bHQET} \nonumber\\
&+ m_Q^2\mathcal{J}\, \left(m_Q\frac{d}{dm_Q}\Phi_{\rm bHQET}\right).
\label{eq:deriv}
\end{align}
Because the function $\Phi_{\rm bHQET}$ on the right-hand side depends on $m_Q$ only through the combination $\omega_i = x_i m_Q$, we have
\begin{align}
& m_Q\frac{d}{dm_Q} \Phi_{\rm bHQET}(x_1 m_Q, x_2 m_Q) \nonumber\\
=& \sum_{i=1,2} x_i \frac{\partial}{\partial x_i} \Phi_{\rm QCD}(x_1,x_2) \,.
\label{eq:scaling}
\end{align}
Substituting Eq.\,\eqref{eq:scaling} into Eq.\,\eqref{eq:deriv} and using Eq.\,\eqref{eq:FacFormula}, we obtain a first-order partial differential equation for $\Phi_{\rm QCD}$:
\begin{align}
&\left( m_Q \frac{\partial}{\partial m_Q} - \sum_{i=1,2} x_i \frac{\partial}{\partial x_i} \right) \Phi_{\rm QCD}  \nonumber\\
& = \big[2 + \gamma(m_Q,\mu)\big] \Phi_{\rm QCD} \,,
\label{eq:massRGE}
\end{align}
where the anomalous dimension $\gamma$ is defined as
\begin{align}
\gamma(m_Q,\mu) \equiv \frac{d}{d\ln m_Q} \ln \mathcal{J}(m_Q,\mu) \,.
\label{eq:gamma_def}
\end{align}
Eq.\,\eqref{eq:massRGE} is the mass evolution equation for the QCD LCDA of $\Lambda_Q$. It can be solved by the method of characteristics. Introducing an auxiliary parameter $s$, we obtain three equations:
\begin{align}
&\frac{dm_{Q}}{ds}=m_{Q} \,,\ \ \ \frac{dx_{i}}{ds}=-x_{i} \,,\nonumber\\
&\frac{d\Phi_{{\rm QCD}}}{ds}=(2+\gamma)\Phi_{{\rm QCD}} \,.
\end{align}
The first two equations imply that $m_Q x_i$ is a constant as  $m_Q$ varies. The first and the third equations lead to
\begin{align}
\frac{d\Phi_{\rm QCD}}{\Phi_{\rm QCD}} = \frac{dm_Q}{m_Q} \big[2 + \gamma(m_Q,\mu)\big] \,.
\end{align}
Integrating from the initial values $m_Q^{(0)},\, x_i^{(0)}$ to $m_Q,\, x_i$ yields the solution
\begin{align}
&\Phi_{\rm QCD}(x_i, m_Q;\mu)\nonumber\\
=&  \left(\frac{m_Q}{m_Q^{(0)}}\right)^2 \Phi_{\rm QCD}\!\left( \frac{m_Q}{m_Q^{(0)}} x_i, m_Q^{(0)};\mu \right)\nonumber\\
&\times \exp\!\left[ \int_{m_Q^{(0)}}^{m_Q} \frac{dm'}{m'} \gamma(m',\mu) \right] \,,
\label{eq:solution}
\end{align}
where the rescaling $x_i^{(0)} \to (m_Q/m_Q^{(0)})x_i$ has been used because $m_Q x_i$ is constant.
Physically, this rescaling reflects the shift of the light-quark momentum fractions, while the exponential factor accounts for the evolution of the jet function.

In order to evaluate the exponential factor, one needs to calculate the anomalous dimension $\gamma$. One should first resum the large logarithms $\ln(\mu/m_Q)$ in the matching kernel $\mathcal{J}$. From the one-loop result in Eq.\,\eqref{eq:Jfunction}, the $\mu$ evolution of $\mathcal{J}$ is
\begin{equation}
\frac{d}{d\ln\mu} \ln\mathcal{J}(m_Q,\mu) = \frac{\alpha_s(\mu)C_F}{\pi}\left(\ln\frac{\mu}{m_Q} + \frac{5}{4}\right) + \mathcal{O}(\alpha_s^2) \,,
\label{eq:J_rge}
\end{equation}
and the initial condition reads
\begin{equation}
\mathcal{J}(m_Q,m_Q) = 1 + \frac{\alpha_s C_F}{\pi}\Big(\frac{\pi^2}{48} + \frac{5}{4}\Big) \,. \label{eq:initialCondition}
\end{equation}
Note that we use the $\overline{\rm MS}$ scheme for $m_Q(\mu)$, whose $\mu$ dependence contributes to the evolution in Eq.\,\eqref{eq:J_rge} only at ${\cal O}(\alpha_s^2)$ and can thus be neglected in the following analysis.
Integrating Eq.\,\eqref{eq:J_rge} with the initial condition in Eq.\,\eqref{eq:initialCondition} gives the resummed matching kernel
\begin{align}
&\mathcal{J}(m_Q,\mu)\phantom{\Big]} \nonumber\\
=& \exp\!\left[ \int_{m_Q}^{\mu} \frac{d\mu'}{\mu'} \frac{\alpha_s(\mu')C_F}{\pi} \Big( \ln\frac{\mu'}{m_Q} + \frac{5}{4} \Big) \right] \nonumber\\
&\times \mathcal{J}(m_Q,m_Q) \,.
\label{eq:J_resum}
\end{align}
Then, using Eq.\,\eqref{eq:J_resum} together with Eq.\,\eqref{eq:gamma_def} and the running coupling equation
\begin{equation}
\frac{d\alpha_{s}(\mu)}{d{\rm ln}\mu}=\beta[\alpha_{s}(\mu)]=-\frac{\alpha_{s}^{2}(\mu)}{2\pi}\beta_{0} \,,\label{eq:alphasEvolu}
\end{equation}
and keeping terms up to $\mathcal{O}(\alpha_s^2)$, one finds
\begin{align}
&\gamma(m_Q,\mu) \nonumber\\
= & -\frac{5}{4}\frac{\alpha_s(m_Q)C_F}{\pi} + \frac{2C_F}{\beta_0} \ln\frac{\alpha_s(\mu)}{\alpha_s(m_Q)} + \mathcal{O}(\alpha_s^2) \,,
\label{eq:gamma_LL}
\end{align}
where $\beta_0 = \frac{11}{3}C_A - \frac{4}{3} n_f T_F$. The logarithm of the $\alpha_s$ ratio arises from the running of $\alpha_s$ between $m_Q$ and $\mu$. Substituting Eq.\,\eqref{eq:gamma_LL} into the exponent of Eq.\,\eqref{eq:solution} and performing the integrals yields
\begin{widetext}
\begin{align}
\Phi_{{\rm QCD}}(m_{Q},x_{i};\mu)= & \left(\frac{m_{Q}}{m_{Q}^{(0)}}\right)^{2}\Phi_{{\rm QCD}}\Big(m_{Q}^{(0)},\frac{m_{Q}}{m_{Q}^{(0)}}x_{i},\mu\Big)\nonumber \\
 & \times{\rm exp}\left\{ \frac{5C_{F}}{2\beta_{0}}{\rm ln}\frac{\alpha_{s}(m_{Q})}{\alpha_{s}(m_{Q}^{(0)})}+\frac{4\pi C_{F}}{\beta_{0}^{2}}\left[\frac{1}{\alpha_{s}(m_{Q})}\ln\frac{\alpha_{s}(\mu)}{\alpha_{s}(m_{Q})\,e}-\frac{1}{\alpha_{s}(m_{Q}^{(0)})}\ln\frac{\alpha_{s}(\mu)}{\alpha_{s}(m_{Q}^{(0)})\,e}\right]\right\} \,.\label{eq:exp_factor}
\end{align}
\end{widetext}

Eqs.\,\eqref{eq:solution} and \eqref{eq:exp_factor} constitute the leading-power mass evolution of the $\Lambda_Q$ QCD LCDA. They allow one to predict the LCDA at any mass $m_Q$ given the initial condition at $m_Q^{(0)}$. Note that the mass evolution in Eq.\,\eqref{eq:exp_factor} has the same form as that in Ref.\,\cite{Wang:2024wwa}, except that the factor $m_Q/m_Q^{(0)}$ is now squared, and the coefficient of the logarithmic term differs. These are the differences between the baryon and meson LCDA cases.

\section{Renormalon Corrections}
\label{sec:power}

The leading-power factorization formula in Eq.\,\eqref{eq:FacFormula} can be modified by incorporating corrections from a renormalon model, which arise from the resummation of bubble-chain diagrams dressing the gluon propagator in virtual exchanges. 
In Ref.~\cite{Guo:2025obm}, such renormalon corrections were evaluated for the heavy-meson LCDA in the peak region, where they are effectively accounted for by a shift in the jet function: $J_{\rm peak}\to J_{\rm peak}+\delta J_{\rm peak}$. 
Since the jet function originates solely from a gluon exchange between the heavy quark and its Wilson line, it is identical for heavy mesons and heavy baryons. 
Therefore, the corresponding bubble-chain diagrams, and consequently the renormalon ambiguity $\delta J_{\rm peak}$, must also be the same for the $\Lambda_Q$ baryon. 
Its explicit expression reads~\cite{Guo:2025obm}
\begin{align}
&\delta J_{\rm peak}(m_Q^2)  \nonumber\\
=& \frac{1}{\beta_0}\, {\mathrm{Res}} \left| \frac{2C_F(1+w)\Lambda_{\rm QCD}^{2w}\Gamma(-2w)\Gamma(w)}{m_Q^{2w}\Gamma(2-w)} e^{5w/3} \right| \,,
\end{align} 
where $w$ is the Borel parameter, and $\delta J_{\rm peak}$ is determined by the residues of the poles in the Borel plane. The leading pole at $w=1/2$ gives the $1/m_Q$ term, and the next pole at $w=1$ yields the $1/m_Q^2$ contribution. Evaluating the residues one obtains the expansion
\begin{align}
\delta J_{\rm peak}(m_Q^2) = & \frac{3C_F \Lambda_{\rm QCD}}{\beta_0 m_Q} e^{5/6} + \frac{C_F \Lambda_{\rm QCD}^2}{\beta_0 m_Q^2} e^{5/3} + \dots \nonumber\\
\equiv & \frac{\Lambda_1}{m_Q} + \frac{\Lambda_2}{m_Q^2} +{\cal O}\Big(\frac{\Lambda_3}{m_Q^3}\Big) \,.
\label{eq:deltaJ}
\end{align}
As argued in Ref.~\cite{Beneke:1998ui}, the renormalon model can be used to 
estimate the magnitude of power corrections in a general factorization framework. 
In this work, we use the renormalon correction $\delta J_{\rm peak}$ to estimate the magnitude of the missing power corrections to the leading-power factorization formula in Eq.\,\eqref{eq:FacFormula}. In particular, we focus on their contribution to the $m_Q$ dependence of $\Phi_{\rm QCD}$. 

It should be emphasized that a full treatment of power corrections would introduce operator mixing and spoil 
the simple product form of the factorization formula in Eq.~\eqref{eq:FacFormula}. 
By contrast, in the renormalon model this product structure remains intact, so 
the differential equation that governs the $m_Q$ dependence of the QCD LCDA can 
still be derived. 
Consequently, the renormalon model does not provide the exact analytic form of these 
power corrections. Instead, it yields a numerical estimate of their magnitude through 
the additive modification $\delta J_{\rm peak}$ in the matching kernel.

We incorporate the bubble-chain correction derived from the renormalon model into the factorization formula and use it to estimate the uncertainty  due to missing power corrections in the leading‑power mass dependence. Including the renormalon correction given in Eq.~(\ref{eq:deltaJ}), the full matching factor becomes
\begin{align}
\mathcal{J}_{\rm corr}(m_Q,\mu) = \mathcal{J}(m_Q,\mu) + \delta\mathcal{J}(m_Q,\mu) \,, 
\label{eq:Jcorr_def}
\end{align}
where $\delta\mathcal{J} = (\bar{f}_{\Lambda_Q}/f_{\Lambda_Q}) \delta J_{\rm peak}$. Expanding $\delta\mathcal{J}$ to $\mathcal{O}(\alpha_s \Lambda_1/m_Q)$ yields
\begin{align}
& \delta\mathcal{J}(m_Q,\mu)\nonumber\\
 = &\frac{\Lambda_1}{m_Q} + \frac{\Lambda_2}{m_Q^2} + \frac{\alpha_s(\mu)C_F}{4\pi}\Big(2\ln\frac{\mu^2}{m_Q^2}+3\Big)\frac{\Lambda_1}{m_Q} \nonumber\\
 &+ {\cal O}\Big(\frac{\alpha_{s}\Lambda_{2}}{m_{Q}^{2}}\Big)+{\cal O}\Big(\frac{\Lambda_{3}}{m_{Q}^{3}}\Big)+{\cal O}(\alpha_s^2) \,.
\end{align}
Here we treat the $\alpha_s \Lambda_1/m_Q$ term as being of the same order as $\Lambda_2/m_Q^2$, and neglect other higher‑order terms such as $\alpha_s^2$ and $\alpha_s \Lambda_2/m_Q^2$.
Accordingly, the correction to the boundary condition is
\begin{align}
\delta{\cal J}(m_{Q},m_{Q})=\frac{\Lambda_{1}}{m_{Q}}+\frac{\Lambda_{2}}{m_{Q}^{2}}+\frac{\alpha_{s}(m_{Q})C_{F}}{\pi}\frac{3\Lambda_{1}}{4m_{Q}}.\label{eq:initialcorrected}
\end{align}
Up to $\mathcal{O}(\alpha_s)$, the $\mu$ evolution of ${\cal J}_{{\rm corr}}^{(1)}(m_{Q},\mu)$ can be obtained by taking the derivative of Eq.~\eqref{eq:Jcorr_def} and using the one-loop result ${\cal J}^{(1)}$ and $\delta{\cal J}$:
\begin{align}
&\frac{d}{d{\rm ln}\mu}{\rm ln}{\cal J}_{{\rm corr}}^{(1)}(m_{Q},\mu) \nonumber\\
=&\left[1-\frac{\delta{\cal J}}{{\cal J}^{(1)}}\right]\frac{1}{{\cal J}^{(1)}}\frac{d}{d{\rm ln}\mu}{\cal J}^{(1)}+\frac{1}{{\cal J}^{(1)}}\frac{d}{d{\rm ln}\mu}\delta{\cal J}\nonumber\\
=&\frac{\alpha_{s}(\mu)C_{F}}{\pi}\left[\Big({\rm ln}\frac{\mu}{m_{Q}}+\frac{5}{4}\Big)-\frac{\Lambda_{1}}{m_{Q}}\Big(\frac{1}{4}+{\rm ln}\frac{\mu}{m_{Q}}\Big)\right] \,.\label{eq:dJcorrdlnmu}
\end{align}
Using the boundary condition in Eq.\,\eqref{eq:initialCondition} and the renormalon-corrected boundary condition in Eq.\,\eqref{eq:initialcorrected}, and summing the large logarithms, one obtains
\begin{align}
&{\cal J}_{{\rm corr}}(m_{Q},\mu)=  {\cal J}(m_{Q},\mu)\Big[1+\frac{\delta{\cal J}(m_{Q},m_{Q})}{{\cal J}(m_{Q},m_{Q})}\Big]\nonumber\\
&\times{\rm exp}\left[-\frac{\Lambda_{1}}{m_{Q}}\int_{m_{Q}}^{\mu}\frac{d\mu^{\prime}}{\mu^{\prime}}\frac{\alpha_{s}(\mu^{\prime})C_{F}}{\pi}\Big(\frac{1}{4}+{\rm ln}\frac{\mu^{\prime}}{m_{Q}}\Big)\right] \,,
\end{align}
where ${\cal J}(m_{Q},\mu)$ is the resummed kernel given in Eq.\,\eqref{eq:J_resum}, and the remaining factor stems from the renormalon corrections. The corresponding anomalous dimension is modified as well, and can be written as 
\begin{align}
\gamma_{\rm corr}(m_{Q},\mu) =& \frac{d}{d\ln m_Q}\ln\mathcal{J}_{\rm corr}(m_{Q},\mu) \nonumber\\
\equiv& \gamma (m_{Q},\mu) + \delta\gamma (m_{Q},\mu) \,,
\end{align}
where the renormalon correction to the anomalous dimension $\delta\gamma(m_Q,\mu)$ reads
\begin{widetext}
\begin{align}
\delta\gamma(m_Q,\mu) =& \frac{\Lambda_1}{m_Q} \frac{C_F}{\beta_0} \Bigg[ 2\ln\frac{\mu}{m_Q} - 2\ln\frac{m_Q}{\Lambda_{\rm QCD}} \ln\frac{\ln(\mu/\Lambda_{\rm QCD})}{\ln(m_Q/\Lambda_{\rm QCD})}  - \frac{5}{2} \ln\frac{\alpha_s(\mu)}{\alpha_s(m_Q)} \Bigg]  \nonumber\\
&+ \frac{\alpha_s(m_Q)C_F}{\pi} \frac{\Lambda_1}{m_Q} \left( \frac{\pi^2}{48} + \frac{3}{4} \right) - \frac{\Lambda_1}{m_Q} - \frac{2\Lambda_2}{m_Q^2} \,.
\label{eq:deltaGamma}
\end{align}
\end{widetext}
The correction $\delta\gamma$ is suppressed by at least one power of $1/m_Q$ and is therefore most relevant for masses close to the charm scale. The complete solution of the mass dependence including renormalon corrections is
\begin{align}
&\Phi_{\rm QCD}(x_i, m_Q;\mu) \nonumber\\
=& \left(\frac{m_Q}{m_Q^{(0)}}\right)^2 \Phi_{\rm QCD}\!\left( \frac{m_Q}{m_Q^{(0)}} x_i, m_Q^{(0)};\mu \right) \nonumber\\
&\quad \times \exp\!\left[ \int_{m_Q^{(0)}}^{m_Q} \frac{dm'}{m'} \left(\gamma(m',\mu)+\delta\gamma(m',\mu)\right) \right] \,.
\label{eq:solution_corr}
\end{align}
It can be seen the renormalon corrections appear as an overall multiplicative factor
\begin{align}
\exp\!\left[ \int_{m_Q^{(0)}}^{m_Q} \frac{dm'}{m'} \delta\gamma(m',\mu) \right] \,.
\end{align}
 on the solution in Eq.\,\eqref{eq:solution}. As a result, they only modify the height of the QCD LCDA in the peak region, causing no horizontal shift.

\section{Numerical analysis}
\label{sec:num}

\begin{figure*}
\centering
\includegraphics[width=1.0\textwidth]{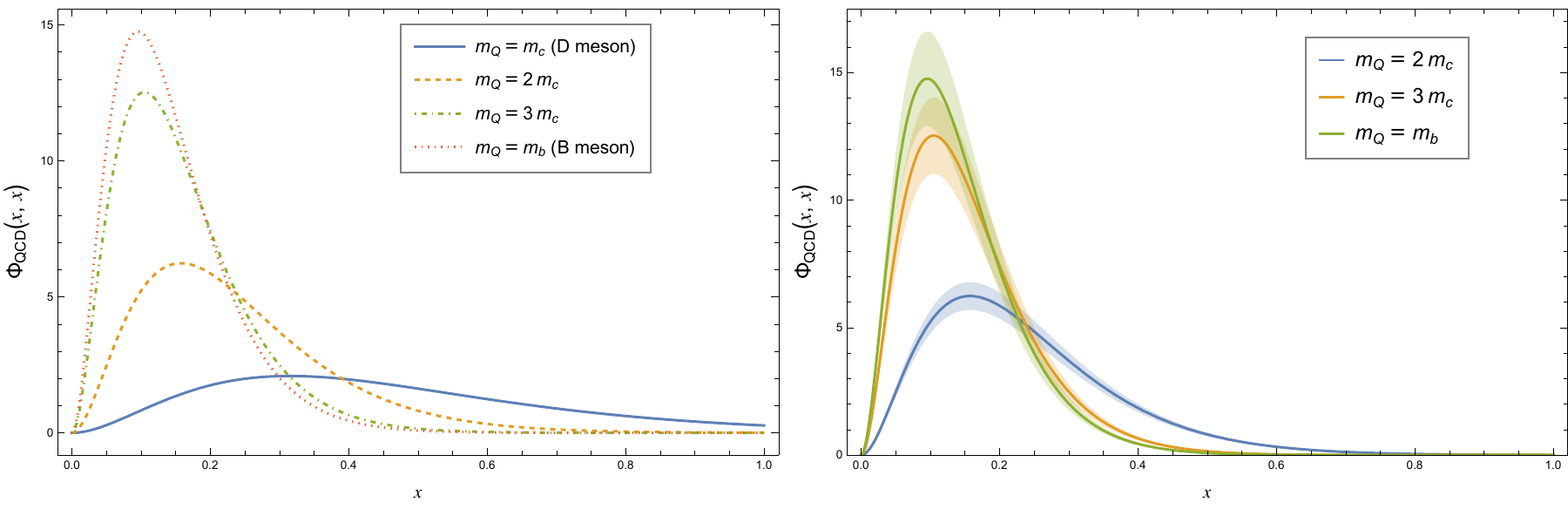}
\caption{Left: Leading-power mass evolution of the $\Lambda_Q$ LCDA in QCD along the symmetric line $x_1=x_2=x$ for different heavy-quark masses. 
Right: The same evolution including power-correction uncertainties from the renormalon model, indicated by the bands.}
\label{fig:phi_combined}
\end{figure*}
\begin{figure*}
\centering
\includegraphics[width=0.6\textwidth]{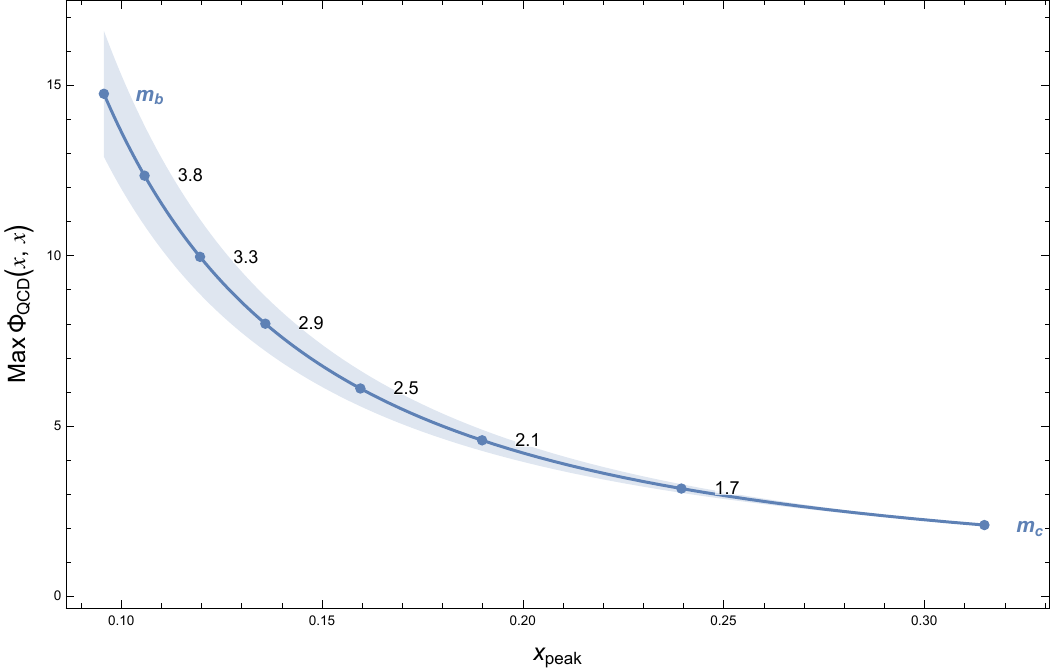}
\caption{Parametric curve of the peak position $x_{\rm peak}$ and the maximum value $\max\Phi_{\rm QCD}$ driven by the heavy‑quark mass $m_Q$. Each point on the curve corresponds to a fixed value of the heavy quark mass $m_Q$. The eight markers on the curve indicate selected values of $m_Q$ between $m_c$ and $m_b$ (in GeV). The shaded band represents the estimated uncertainty from the renormalon correction and serves as a measure of the size of the missing power corrections.}
\label{fig:Moments}
\end{figure*}
To illustrate the mass dependence of the $\Lambda_Q$ QCD LCDA, we adopt the simple exponential model for the $\Lambda_Q$ HQET LCDA~\cite{Ball:2008fw}
\begin{align}
\Phi_{\rm bHQET}(\omega_1,\omega_2) = \frac{\omega_1\omega_2}{\omega_0^4} e^{-(\omega_1+\omega_2)/\omega_0} \,,
\label{eq:model}
\end{align}
with $\omega_0 = 0.4$ GeV. We first obtain the initial QCD LCDA $\Phi_{\rm QCD}(x_i, m_c; \mu)$ from the matching formula at $\mu = 8$ GeV with $m_Q = m_c = 1.27$ GeV. Taking this distribution as the initial condition at $m_Q^{(0)} = m_c$, we evolve the distribution to higher masses up to $m_Q = m_b = 4.18$ GeV using Eq.\,\eqref{eq:solution_corr} and Eq.\,\eqref{eq:exp_factor}, both with and without the inclusion of the renormalon correction $\delta\gamma$. All quark masses are defined in the $\overline{\rm MS}$ scheme. The  parameters in Eq.\,\ref{eq:deltaJ}) evaluate to $\Lambda_1 = 0.332$ GeV and $\Lambda_2 = 0.076$ GeV$^2$ for $\Lambda_{\rm QCD}=0.3$ GeV.

The left panel of Fig.\,\ref{fig:phi_combined} displays the LCDA projected onto the symmetric line $x_1 = x_2 \equiv x$ for four values of $m_Q$:  $m_c$, $2m_c$, $3m_c$, and $m_b$ without renormalon corrections. As $m_Q$ increases, the distribution becomes significantly narrower and its peak height grows, while the peak position shifts towards smaller $x$. This behavior is a direct consequence of the rescaling $x_i \to (m_Q/m_Q^{(0)}) x_i$ together with the overall multiplicative factor $m_Q/m_Q^{(0)}$ appearing in Eq.\,\eqref{eq:exp_factor}. The right panel of Fig.\,\ref{fig:phi_combined} shows the same evolution with the renormalon correction included, where the bands represent the uncertainty introduced by the modification $\delta\gamma$ in Eq.\,\eqref{eq:solution_corr}.  We estimate the magnitude of the power corrections from the band width. The bands are observed to become slightly wider with increasing $m_Q$. This does not contradict the expectation that the power corrections themselves decrease with $m_Q$, because in Eq.\,\eqref{eq:solution_corr} the power‑correction exponential depends on the integral of $\delta\gamma$ from $m_Q^{(0)}$ to $m_Q$. As $m_Q$ grows, the integrated range expands, which can overcome the $1/m_Q$ suppression of $\delta\gamma$. It should be noted that the mass dependence derived in this work is valid only in the peak region. Therefore, only the behavior in the small‑$x$ region of Fig.\,\ref{fig:phi_combined} is physically meaningful, while the tail region is not the focus of this work, so we do not pursue it further here. 
 
Fig.\,\ref{fig:Moments} displays the relation between the peak position $x_{\rm peak}$ and the maximal value $\rm{Max}~\Phi_{\rm QCD}$ as the heavy-quark mass $m_Q$ varies. Each point on the curve corresponds to a fixed value of the heavy quark mass $m_Q$. The markers on the curve indicate selected values of $m_Q$ between $m_c$ and $m_b$ (in GeV). The shaded band represents the estimated uncertainty from the renormalon correction and serves as a measure of the size of the missing power corrections.

\section{Conclusions}
\label{sec:concl}
We have studied the heavy-quark mass dependence of the leading-twist QCD LCDA of the $\Lambda_Q$ baryon. 
Based on the factorization between the QCD and boosted HQET LCDAs in the peak region, we have derived a first-order partial differential equation governing this mass dependence at a fixed renormalization scale. 
The equation has been solved analytically, revealing a simple rescaling of the light-quark momentum fractions together with an exponential factor arising from the jet function.

To estimate contributions that are suppressed by $\Lambda_{\rm QCD}/m_Q$, we incorporated Borel‑resummed perturbative corrections from a renormalon model. 
Specifically, we used the renormalon correction $\delta J_{\rm peak}$ to estimate the magnitude of the missing power corrections to the leading-power factorization formula in Eq.\,\eqref{eq:FacFormula}, focusing on their contribution to the $m_Q$ dependence of $\Phi_{\rm QCD}$. 
These corrections are encoded in a modified jet function and an additional term in the mass anomalous dimension. 
The complete solution including these corrections has been derived and used in a numerical analysis. 
Using an exponential model for the HQET LCDA, we have studied how the LCDA varies with $m_Q$ from the charm mass up to the bottom mass. 
The distribution gradually narrows and its peak shifts toward smaller $x$ as $m_Q$ increases. 
The uncertainties induced by the renormalon corrections are moderate and remain under control.

The mass-dependence formalism developed in this work is an essential ingredient for the systematic determination of the $\Lambda_b$ LCDA from lattice QCD. 
It enables a reliable extrapolation of results obtained at computationally accessible heavy-quark masses to the physical $b$-quark scale. 
Together with the existing two-step matching procedure, our results open the door to precision predictions for exclusive heavy-baryon decays within factorization-based approaches.

\section*{Acknowledgements}
We thank Jia-Lu Zhang for valuable discussions. This work is supported in part by the National Natural Science Foundation of China under Grants Nos. 12305103, 12475098, and 12105247. J.X is supported in part by the Key Laboratory for Particle Astrophysics and Cosmology, Ministry of Education (MoE).

\end{document}